\documentstyle[preprint,aps,epsfig]{revtex}

\tighten
\begin{document}
\draft
\title{The Average Kinetic Energy of the Superconducting State}

\author{ Mauro M. Doria$^{a,*}$,  S. Salem-Sugui Jr.$^{a}$,  I. G. de Oliveira$^{b}$, L. Ghivelder$^{a}$, and E. H. Brandt$^{c}$}
%\footnote{$^*$Corresponding author, e-mail: mmd@if.ufrj.br}
\address{ $^a$ Instituto de F\'{\i}sica, Universidade Federal do 
Rio de Janeiro,\\ C.P. 68528, 21945-970, Rio de Janeiro RJ, Brazil\\ 
$^b$ Faculdade de Ci\^encias Exatas e Tecnol\'ogicas, Universidade Igua\c{c}u - UNIG, Nova Igua\c cu, 26260-100 RJ, Brazil, and \\
$^c$ Max Planck Institute f\"ur Metallforschung, Intitut f\"ur Physik, D-70506 Stuttgart, Germany}

\date{\today}

\maketitle

\begin{abstract}
Isothermal magnetization curves are plotted as the magnetization times the magnetic induction, $4 \pi M \cdot B$, versus the applied field, $H$. 
We show here that this new curve is the average kinetic energy of the superconducting state versus the applied field, for type-II superconductors with a high Ginzburg-Landau parameter $\kappa$.
The maximum of $4 \pi M \cdot B$ occurs at a field, $H^{*}$, directly related to the upper critical field, $H_{c2}$, suggesting that $H_{c2}(T)$ may be extracted from such plots even in cases when it is too high for direct measurement.
We obtain these plots both theoretically, from the Ginzburg-Landau theory,  and experimentally, using a  Niobium sample with $T_c = 8.5\; $K, and compare them.
\end{abstract}

%\begin{multicols}{2}
\narrowtext
%\widetext

\section{Introduction} 
\label{sec.int}

The average kinetic energy  in Classical Physics is a useful quantity that helps to obtain important information about several many-body systems, such as the Sun, whose  internal temperature can be estimated in this way\cite{Berkeley}.
In cases when the interaction among particles is only partly known, and the Virial Theorem applies,  the total energy can be directly related to the average kinetic energy, and the average kinetic energy to  the  equipartion theorem, leading to the prediction of  thermal properties of the classical system.
In this paper we demonstrate, both theoretically and experimentally, that the average kinetic energy of the superconducting state can be directly read from isothermal magnetization curves, for superconductors such that the Ginzburg-Landau parameter $\kappa$ is larger than a few units. 
Thus this average only involves the paired electrons that form the condensate.

It is often the case in Physics that just expressing  data in a different way is sufficient to bring a new insight into the problem under investigation.
This is exactly the present situation and  we claim here that an interesting and new
insight  is obtained if isothermal magnetization ($4 \pi M$) data is plotted differently, namely, as the product $4 \pi M \cdot B$, where the magnetic induction is $B=H+4\pi M$ and $H$ is the equilibrium applied field.
In this paper we show that this new curve $4 \pi M \cdot B$ directly determines the average kinetic energy for superconductors with  $\kappa > \kappa_c$,  with an accuracy of less than one percent for  $\kappa_c \approx 3$.

A remarkable property of the product $4 \pi M \cdot B$ is that it may have several local minima along the mixed state.
This property  follows from the fact that  $4 \pi M \cdot B$  vanishes at both critical fields,
$H_{c1}$ and $H_{c2}$. 
At $H_{c1}$, $M$ reaches a minimum  and $B$ vanishes, whereas at 
$H_{c2}$, $B$ is maximum and $M$ vanishes. 
Since $4 \pi M \cdot B$ vanishes at two fields, the two extremes of the mixed state, it must necessarily have an absolute  minimum in between, and possibly several local minima.
The connection between the average kinetic energy and the isothermal magnetization follows from  the Virial Theorem for superconductivity\cite{DGR} which states that,
\begin{eqnarray}
{{{\bf H}\cdot{\bf B}}\over{4\pi}} = F_{kin} + 2 \;F_{field}, \label{virial}
\end{eqnarray}
where $F_{kin}$ and  $F_{field}= \langle{\bf h}^2/8\pi \rangle$, are  the average kinetic energy and the  average magnetic energy, respectively. 
The local magnetic field inside the superconductor is ${\bf h}$ and $\langle \cdots \rangle$ 
represents an average value  taken over the volume of the superconductor.
This theorem has been discovered rather late for the Ginzburg-Landau (GL) theory and 
it has been proven very useful since then, because it avoids taking the derivative of the free energy for the determination of $H$. The theorem has been extended to the microscopic theory in the context of the quasi-classical Eilenberger theory\cite{KP}.
There it was found that the above theorem holds for temperatures well below $T_c$ and also in presence of non-magnetic impurities.
Thus one expects that even in presence of some types of pinning the above relationship still holds.

In this paper we obtain $4 \pi M \cdot B$ curves both theoretically and experimentally for
isotropic superconductors. We only discuss here examples such that the symmetry along the direction defined by the applied field turns the theoretical problem into a two-dimensional one.
In this case the vector notation is no longer necessary and so, is dropped hereafter.
The $4 \pi M \cdot B$ theoretical curve is derived in three situations of the macroscopic GL theory.
We obtain analytical expressions for it in the Abrikosov and the London limits. we also calculate this curve numerically, using the iterative scheme proposed by E.H. Brandt\cite{Brandt1}, a very accurate method that gives the periodic solution of the GL theory for any field in the mixed state.
Using the iterative method we show that for $\kappa$ larger than a few units the curve $4 \pi M \cdot B$ indeed gives the average kinetic energy.
The absolute minimum of the new curve, 
\begin{eqnarray}
{{d(4 \pi M \cdot B)}\over{d H}}|_{\displaystyle H^{*}} = 0.
\end{eqnarray}
determines a critical field called $H^{*}$. 
A consequence of the connection established here is that this field determines the maximum average kinetic energy, and from this one obtains  the maximum root mean square current density, $J_{max} \equiv \sqrt{\langle J^2 \rangle}$. 
Thus using the present approach one obtains information about the maximum current density without the use of Bean's model. For this one needs the London expression $F_{kin} \approx 2 \pi \lambda^2 \langle J^2 \rangle /c^2$, where $\lambda$ is the penetration depth.
Actually at $H^{*}$ the order parameter is not constant but we  consider as such, to obtain the order of magnititude of this maximum current density circulating in the superconducting state.

In this paper we also show that the ratio $H^{*}/H_{c2}$ becomes $\kappa$ independent
for $\kappa$ larger than a few units.
This property can be useful because it suggests  that $H_{c2}$ can be determined at a lower field, namely, $H^{*}$.
Frequently the upper critical field of many compounds cannot be reached experimentally just because it falls beyond the capabilities of a given experimental set up.

The $4 \pi M \cdot B$ experimental curves were obtained here using a Niobium sample with an approximately spherical shape, mass $m = 0.6487 \; g$, and critical temperature $T_c = 8.5\; $K with a transition width of $\Delta T \sim 0.3 $K. The sample was obtained in an arc melt furnace from 99.9 \% pureness Nb wire. X-ray diffraction done in this sample shows the expected metallic Nb phase. Isothermal magnetization data was always obtained starting from a zero-field cooled procedure and using a commercial  Quantum Design PPMS extraction magnetometer facility.

New features and properties, not visible in the traditional way of plotting  magnetization curves, are unveiled here by this $4 \pi M \cdot B$ plot. Some of them are explained in this paper, but there are others, such as the multi-minima structure in the irreversible region, that require further discussion and is not done here. In the reversible region, only one minimum remains and we find good agreement between theory and experiment. Close to $T_c$ the experimental and theoretical $4 \pi M \cdot B$ curves show good agreement and  the experimental ratio $H^{*}/H_{c2}$ is well predicted by the present theoretical models. Notice that the Virial Theorem, Eq.(\ref{virial}), holds for the  microscopic theory \cite{KP}, and for this reason  the connection between the curve $4 \pi M \cdot B$ and $F_{kin}$, remains valid beyond the GL theoretical framework of this paper.

This paper is organized as follows. In Section \ref{sec.theo} we show that the curve $4 \pi M \cdot B$ gives the average kinetic energy of the superconducting state and  derive this curve both analytically and numerically in different theoretical situations. 
The field  $H^{*}$ is calculated and, its properties, discussed here.
In Section \ref{sec.exp} the experimental $4 \pi M \cdot B$ curves for the Nb sample are obtained, and  the field  $H^{*}$ 
determined. The phase diagram of this Nb sample is obtained with the curve $H^{*}(T)$ included.
Finally we draw conclusions in Section \ref{sec.concl}.

\section{Theory}
\label{sec.theo}

Let us first obtain analytical expressions for the energy function $4 \pi M \cdot B$ in  two  simple situations and determine there the critical field $H^{*}$.
The lower and upper  critical fields are given by $H_{c1} = (\Phi_0/4\pi\lambda^2)\ln{\lambda/\xi}$ and $H_{c2} = \Phi_0/2\pi\xi^2$, respectively,
$\xi$ is the coherence length and  $\kappa=\lambda/\xi$ the Ginzburg-Landau parameter.
The first case investigated here is Abrikosov's theory which calculates the vortex lattice near $H_{c2}$  in the framework of the GL free energy.
There\cite{Tinkham} the magnetic induction is given by
\begin{eqnarray}
B = H - {{H_{c2}-H}\over{(2\kappa^2-1)\beta_{A}}},
\end{eqnarray}
where $\beta_A =  \langle|\Delta|^4 \rangle/\langle|\Delta|^2\rangle^2$ is a parameter determined from  the vortex lattice  symmetry, $\Delta$ being the wavefunction that describes the superconducting state.
One obtains that
\begin{eqnarray}
4 \pi M \cdot B =  - {{H_{c2}-H}\over{(2\kappa^2-1)\beta_{A}}}\big( H - {{H_{c2}-H}\over{(2\kappa^2-1)\beta_{A}}}\big),
\end{eqnarray}
is a parabola that vanishes at $H_{c2}/(1+(2\kappa^2-1)\beta_{A})$ and at $H_{c2}$. 
Although Abrikosov's theory is valid only near $H_{c2}$, still, for $\kappa \gg \sqrt{2}/2$, the first vanishing field  approaches
$2/(1+2\beta_{A})\Phi_0/4\pi \lambda^2$, which is of the same order of magnitude as the true
$H_{c1}$ field.
The absolute minimum of $4 \pi M \cdot B$ is achieved at the field
\begin{eqnarray}
H^{*} =  {{H_{c2}}\over{2}} \big( {{2\beta_A \kappa^2 -\beta_A +2}\over{2\beta_A\kappa^2-\beta_A+1}}\big).
\label{H*abrikosov}
\end{eqnarray}
A simpler expression is obtained noting that $\beta_A$ is very close to one, which is true for
the triangular lattice ($\beta_A = 1.1595953\cdots$): $ H^{*} \approx  (H_{c2}/2)*(1+1/\kappa^2)$.
Thus for any superconductor  with $\kappa$ larger than a few units, this critical field is given by $H \approx 0.5 \; H_{c2}$.

The second model is London theory, known to provide a good description of the superconducting
state in the intermediate field region far away from the critical fields.
This is a fairly good approximation as long as variations of the order parameter near the vortex cores can be ignored, and so, the order parameter can be considered constant all over the superconductor.
In this region the magnetic induction is very well approximated by\cite{Tinkham}
\begin{eqnarray}
B = H - H_{c1} {{\ln{(H_{c2}/B)}}\over{\ln{\kappa}}}.
\end{eqnarray}
Then the energy function becomes
\begin{eqnarray}
4 \pi M \cdot B =  - B H_{c1} {{\ln{(H_{c2}/B)}}\over{\ln{\kappa}}}.
\end{eqnarray}
whose minimum absolute is at the field value
\begin{eqnarray}
H^{*} =  {{H_{c2}}\over{e}} - {{H_{c1}}\over{\ln{\kappa}}} \label{hlon}.
\end{eqnarray}
Since $H_{c1}/H_{c2} = \ln{\kappa}/2\kappa^2$, notice, once more, that in case  $\kappa$ is
larger than a few units the last term can be neglected, rendering  $H^{*} \approx 0.37 H_{c2}$.

The product $4 \pi M \cdot B$ is related to the average kinetic energy of the condensate, without any approximation, through   the  Virial Theorem (Eq.(\ref{virial})).
\begin{eqnarray}
4 \pi M \cdot B  = - F_{kin} / 2 -  \langle \big( h - \langle  h \rangle \big )^2 \rangle
\label{mb}
\end{eqnarray}
is expressed here in reduced units, such that the thermodynamical fields and energies are scaled in the following way: $ h \rightarrow h/\sqrt{2} H_c$, $ H \rightarrow H/\sqrt{2} H_c$, and $F_{kin} \rightarrow  F_{kin}/(H_c^2/4\pi)$, $H_c$ being the thermodynamic field.
The same scaling holds for the magnetic induction which is 
just the average local field, $ B \equiv \langle h \rangle $.
These reduced units are convenient because to retrieve
the above quantities at any given temperature $T$ from a single temperature independent plot, it is enough to scale energies by $H_c(T)^2/4\pi$, and fields by $\sqrt{2}H_c(T)$.

To show that the function $4 \pi M \cdot B$ yields the average kinetic energy, $F_{kin}$, is just a matter of determining how the mean square deviation of the local field, $\langle( h - \langle h \rangle )^2\rangle$, behaves as a function of $\kappa$.
Thus is  done here for the GL theory using the
iterative method of E. H. Brandt\cite{Brandt1}.
This iterative procedure assumes an initial solution, from which one obtains very accurately, after a few iterative steps, the local magnetic field and the order parameter for any applied field in the mixed state.
The $4 \pi M \cdot B$ (continuous line) and $F_{kin}$ (points) curves, obtained using this method, are shown in Fig.\ \ref{mbh} for several $\kappa$, ranging from $0.75$ to $50$. 
The conclusion is that these two  curves, $4 \pi M \cdot B$ and $F_{kin}/2$ become very similar for large $\kappa$.
The field $H^{*}$ is obtained by determining the absolute minimum  of the $4 \pi M \cdot B$ curves in Fig.\ \ref{mbh}, and Fig.\ \ref{hstark} summarizes these results as a $H^{*}/H_{c2}$ vs $\kappa$ plot.

The difference between $4 \pi M \cdot B$ and $F_{kin}/2$ is best quantified , as function of $\kappa$, by two variance ratios introduced here and both computed at the field $H^{*}$.
We compute $\langle(h-\langle h \rangle)^2 \rangle/|4 \pi M \cdot B|_{H^{*}}$ for 
a few points ($\kappa$,$\; \langle (h-\langle h \rangle)^2 \rangle /|4 \pi M \cdot B|_{H^{*}}$): ($0.75,\; 0.11$), ($0.85,\; 0.094$), ($1.0,\; 0.072$), ($1.5,\; 0.034$), ($2.0,\; 0.020$), ($3.0,\; 0.0088$), ($5.0,\; 0.0032$), ($10.0,\; 7.83 \;10^{-4}$), ($20.0,\; 1.99 \; 10^{-4}$), ($30.0,\; 8.8 \; 10^{-5}$), and ($50.0,\; 3.2 \; 10^{-5}$).
Notice that for $\kappa > 3$ the two curves can be regarded as the same
with an error of less than one per cent in the neighborhood of $H^{*}$.
This  error is not uniform for all fields, being  smaller near $H_{c2}$  and larger near $H_{c1}$. 
In Fig.\ \ref{hstark} one has that  $H^{*}/H_{c2}= 0.51$ for $\kappa=3.0$, and essentially, for $\kappa > 3$,  $H^{*}$ is fairly well approximated by its limiting value of $0.5 \; H_{c2}$.
The inset of Fig.\ \ref{hstark} shows the relative root mean square deviation of
the local magnetic field, $\big(\langle(h-\langle h \rangle )^2\rangle \big)^{1/2}/\langle h \rangle |_{H^{*}}$, as a function of $\kappa$.
Again we see a sharp drop of this quantity as $\kappa$ increases.
For instance for $\kappa=3.0$, $\big( \langle ( h- \langle h \rangle )^2 \rangle \big )^{1/2}/ \langle h \rangle |_{H^{*}} = 0.0213$.

Analytical  expressions for these two variances are obtained in the Abrikosov approximation 
which  starts from the well-known Abrikosov relation between the local field and the order parameter\cite{Brandt2}, $ h = H - 2\pi |\Delta |^2/\kappa $, in reduced units.
From it one obtains an interesting expression that relates thr variance to the magnetization:
$ \langle(h-\langle h \rangle )^2\rangle = (4 \pi M)^2 (\beta_A -1) $.
From it one finds that at the field $H^{*}$, given by Eq.(\ref{H*abrikosov}), the following expressions:

\begin{eqnarray}
\frac{\big(\langle(h-\langle h \rangle )^2\rangle \big)^{1/2}}{\langle h \rangle} |_{H^{*}}
= \big ( \frac{\beta_A-1}{\beta_A}\frac {1}{4\beta_A\kappa^4-4(\beta_A-1)\kappa^2+(\beta_A-1)}\big )^{1/2},
\end{eqnarray}
and,
\begin{eqnarray}
\frac{\langle(h-\langle h \rangle )^2\rangle}{|4 \pi M \cdot B|} |_{H^{*}}
= \frac{\beta_A-1}{2 \beta_A \kappa^2 + 1 - \beta_A }.
\end{eqnarray}

Both expressions display a $1/\kappa^2$ behavior for large $\kappa$.
For the triangular lattice, one obtains that
$\big( \langle ( h- \langle h \rangle )^2 \rangle \big )^{1/2}/ \langle h \rangle  |_{H^{*}} \approx 0.172256/\kappa^2$, and $\langle (h-\langle h \rangle)^2 \rangle /|4 \pi M \cdot B|_{H^{*}} \approx 0.068815/\kappa^2$, which agree quite well with the numerical method results.
We conclude that the curve $4 \pi M \cdot B$ can be 
obtained from  the average  kinetic energy for $\kappa > \kappa_c$, with a precision better than one percent for $\kappa_c \approx 3$. Next we study  a Niobium sample and obtain  the curve $4 \pi M \cdot B$ experimentally, starting from the isothermal magnetization curves.

\section{Experiment}
\label{sec.exp}

To exemplify the present ideas we have studied  superconductor Niobium, which along the years has been  an important reference for the understanding of type II superconducting properties\cite{Ling}. The observation of the irreversibility line in pure Nb \cite{Suenaga} as observed in high-$T_c$ superconductors\cite{Blatter}, has renewed the interest on this superconductor. Also, effects which are easier to study and understand in Nb, such as surface pinning\cite{Mishra}, vortex avalanches \cite{Nowak}, and peak effect\cite{Yakov}, has shed some light on the understanding of similar effects in high-Tc superconductors.

The sample has critical temperature below of pure Nb ($T_c=9.2$K), indicating that it contains non-magnetic impurities.  This  causes no problem here, on the contrary,
the impurities  turn the compound into a more type II superconductor, with a larger value of $\kappa$.
Very pure Nb\cite{Finnemore} has $\kappa \approx 1.0$, whereas we find here $\kappa \approx 4$
for this sample, when comparing the present  theory to the experiment.

Next we discuss our data as shown in the figures.

The transition properties of this sample are seen in Fig.\ \ref{expmt}, which shows
a rather small reversible temperature region.
For fields below $4.0\;$kOe, the magnetization becomes essentially irreversible.
Values of $H_{c2}(T)$ are extracted from this diagram following the standard linear extrapolation of the magnetization curve.

Fig.\ \ref{expmh} shows a total of nine isothermal magnetization cycles, $0.5 \; $K apart from each other, for temperatures ranging from $3.5 \; $K to $8 \; $K.
Notice that in the neighborhood of  zero applied field all the magnetization
curves overlap forming  one single temperature independent line.
The slope of this line defines the demagnetization factor.
For finite size samples the demagnetization factor must be included\cite{Jackson}; for a sphere this factor is $1/3$ such that $B = H_{in} + 4\pi M /3$, where 
$H_{in}= H + 4\pi M/3$ is the internal field.
Our sample is an ellipsoid with axis  $ 2 r_1= 4.7\;$mm and
$2 r_ 2= 5.3\;$mm, whose sphere with equivalent volume has radius $5.1 \;$mm.
We have experimentally determined the demagnetization of our sample using the following procedure. We assume that $B = H + 2d M $ where $d$ is a factor to be determined from our data such that, in the Meissner phase, the magnetic induction must vanish.
We find that, in this region, a linear extrapolation of the magnetization data for all $4 \pi M$ vs $H$ curves produces the same value of $2d = 4.16$, independent of temperature.

Another important issue here is the removal of the background magnetization as a function of $H$ for all  temperatures. 
The construction of the curve $4 \pi M \cdot B$ is only useful, e.g. to determine the critical field $H^{*}$, if  all magnetic contributions that are not of superconducting origin are totally removed from  the original isothermal $M$ vs $H$ data. 
For the  Nb sample the following procedure was applied.
The raw magnetization data displays a linear non-zero positive slope as a function of $H$
that goes beyond the $H_{c2}$ field.
This curve is universal and temperature independent.
We subtract from the raw magnetization data at all temperatures this single linear curve $M$ vs $H$.
This is possible because the background magnetization  is temperature independent, and just a single measurement of the paramagnetic background signal close or above $T_c$ is enough to subtract the unwanted  background for any temperature.
It is well-known that this source of magnetic signal is the Pauli paramagnetism of the normal electrons.

Selected isothermal  curves $2dMB$ vs $H$ are shown for  three different temperatures, corresponding to $T=4.5, \;6,\mbox{and} \; 8 \;$K (Figs.\ \ref{exp2dmbha}, \ref{exp2dmbhb}, and \ref{exp2dmbhc}, respectively).
The basic  features of the $M$ vs $H$ curves, such as the critical fields and the peak effect, are also present in the  $2dMB$ vs $H$ curves, 
although in a somewhat different way.
However the $2dMB$ shows a new feature, not noticed in the traditional $M$ vs $H$ plot,
namely the new   minimum  at  $H^{*}$. 
For $T=4.5\; $K the absolute minimum of the curve $2dMB$ occurs at $H=6.0 \;$kOe, a field  associated with the peak effect. Still the $H^{*}$ is observable there.
However for increasing temperature, $T= 6 \; $K,
$H^{*}$ becomes the absolute minimum and the peak effect turns into a local minimum, in this case around $4 \; $kOe. 
Finally at $T=8$K only one minimum remains, $H^{*}$, and the peak effect is essentially absent.
Notice that  two $H^{*}$ fields are defined, measured in  ascending (zero field cooled - zfc) and  descending  (field cooled - fc) fields.
The zfc and fc  $H^{*}$ fields are separated by approximately $500 \;$Oe for the $T= 4.5 \;$K curve whereas for  $8 \;$K they become the same.

In Fig.\ \ref{comp} we compare the  $\kappa = 3,\;4, \;\mbox{and}\; 5$  theoretical  and the $T = 8.0 \; $K experimental curves obtained in this paper.
All $2dMB$ data sets are normalized along the $2dMB$ axis to their maximum absolute values, achieved at $H^{*}$.
Along the $H$ axis, the theoretical curves are scaled by the inverse of  their upper critical fields, such as in Fig.\ \ref{mbh}.
However for the experimental curve we adopted another normalization procedure, namely we scaled fields by $796.98 \;H$ due to the following reason. 
Our goal is to shift the $2dMB$ experimental curve along the $H$ axis in search of the best theoretical fit, which we find to occur when experimental and theoretical $H^{*}$ fields coincide. The experimental absolute maximum of $|2dMB|$ is reached at  $H^{*} = 329.81\;$Oe. 
Next we apply the scale transformation $\alpha H/329.81$ and search for $\alpha$ that gives  the best agreement with the theoretical curves. This happens at $\alpha = 0.46$, very close to the theoretical expected ratio of $\alpha = H^{*}/H_{c2} \approx 0.5$ of the theoretical curve, described in  Fig.\ \ref{hstark}.
Notice that the three theoretical curves coincide near $H_{c2}$, as they should, because of the normalization, but not near $H_{c1}$.
From this figure we conclude that the theoretical $\kappa = 4$ curve provides the best fitting to the experimental data.
It is worth mentioning that this value of $\kappa$ agrees with its experimental estimate derived from the ratio of the critical fields. $H_{c2}/H_{c1}$.

Fig.\ \ref{expht} presents the phase diagram, obtained from the values of $H_{c2}$ that follows from the curves of Fig.\ \ref{expmt}. 
Also plotted in this figure are the values of $H^{*}$  found in the increasing and decreasing field branches of plots, such as shown in Figs.\ \ref{exp2dmbha}, \ref{exp2dmbhb}, and \ref{exp2dmbhc}. 

We find a quite perfect match between the experimental values of $H^{*}$ and $H_{c2}(T)$ with the expression obtained from the theories previously discussed.
Thus for this Nb sample we find that the curve $H_{c2}/2$ fits fairly well the experimental $H^{*}$ curve. 

We  estimate the maximum root mean square average current, $J_m$, for $T=8\;$K.
Firstly one must obtain the London penetration depth from the 
lower critical field for this temperature: $H_{c1} \approx 18 \;$Oe. We obtain that $\lambda \approx 0.36 \; \mu$m and taking that the 
maximum of  $|2 d M \cdot B|$ is reached at $3.88 \; 10^3 \;\mbox{G}^2$, it follows that $J_m \approx 4.2 \; 10^5\; \mbox{A.cm}^{-2}$. Notice that this maximum mean square average current density is being obtained in the {\it reversible} regime.

In summary we find here that this $4 \pi M \cdot B$ curve  naturally enhances many aspects of traditional $4 \pi M$ vs $H$ plots, such as the peak effect,  the first entry field ($H_{c1}$), which is also much more clearly resolved in this kind of plot.
Obviously the theories considered in the previous section do not describe  the many features of the experimental curves $2dMB$ vs $H$, especially when irreversible processes are present. 
This is expected because those theories do not take into account pinning effects.
When pinning effects become small and there is reversibility in the magnetization,
we found agreement between theory and experiment. 
It is always possible to cast isothermal magnetic data as a $4 \pi M \cdot B$ curve, but the question is whether the present connection between the $4 \pi M \cdot B$ curve and  $F_{kin}$ still holds when irreversible effects are present, such as the peak effect.
We believe that this connection indeed  remains valid in the irreversible region
in some large $\kappa$ limit.
Obviously this connection should always be regarded approximately and corrections  are expected  due to pinning.
We believe that these corrections should be small, based on the fact
that the Virial Theorem gets little, or no modification  at all\cite{KP} when
some types of defects\cite{DGR,DS1,DS2} are present.
For temperatures close to $T_c$, where irreversible  effects become small, 
we found that the present ideas hold  and the average kinetic energy interpretation of
the curve $4 \pi M \cdot B$ is applicable.

\vskip 1.0truecm
\section{Conclusions}
\label{sec.concl}

In conclusion the present proposal of casting isothermal magnetization data as the product of the magnetization times the magnetic induction is useful because it directly gives the average kinetic energy of the superconducting state for any applied field.
We have studied here this curve both theoretically and experimentally.
The field of maximum average kinetic energy in the condensate is just the minimum point of the new curve.
From this field  we estimate the maximum mean square average current density.
We show using the Ginzburg Landau theory that this field value is half the upper critical field value for large $\kappa$ superconductors.
Thus its experimental determination can help to understand properties of the upper critical field which often falls beyond the experimental capabilities of a given equipment.
We found analytical expressions for this curve in two special limits of the Ginzburg-Landau theory, namely the Abrikosov and the London limits.
In the general situation we have  obtained this curve numerically for several values of
the Ginzburg-Landau constant, using an efficient iterative method\cite{Brandt1} to solve the Ginzburg-Landau theory.
We find that for a Ginzburg-Landau constant larger than a few units the mean square deviation
of the local magnetic field becomes extremely small.
Only for such parameter values the product magnetization times the magnetic induction 
gives the average kinetic energy.

In summary we find the present proposal of plotting the isothermal magnetization useful  because it provides a new insight into the properties of the mixed state.
%Numericals
%Geometry

\acknowledgments
This work was supported by a collaboration agreement DAAD-CNPq and in part by FAPERJ
and CAPES.

%\end{document}

%\pagebreak

\begin{center}
\begin{figure}
\caption{ Two theoretical curves, $4 \pi M \cdot B$ (continuous)  and $F_{kin}/2$ (dots), versus $H$ are shown here  for several $\kappa$, namely, $0.75$, $0.85$, $1.0$, $1.2$, $1.5$, $2.0$, $3.0$, and $5.0$. For higher $\kappa$, $10$, $20$, $30$, and $50$, the two curves, $4 \pi M \cdot B$ (continuous) and $F_{kin}/2$ (points) are shown in the inset. The field $H^{*}$, the absolute minimum of the $4 \pi M \cdot B$ curve, is suggested for a few curves, as examples.
The inset clearly shows that for large $\kappa$ $H^{*}/H_{c2}\approx 0.5$.}
\label{mbh}
\end{figure}
\end{center}

\begin{center}
\begin{figure}
\caption{ The field $H^{*}$, the absolute minimum of the $4 \pi M \cdot B$ curve, is plotted here as a function of $\kappa$. 
For $\kappa$ larger than a few units the  ratio $H^{*}/H_{c2}$ approaches $0.5$.
The inset shows that the root mean square deviation of the local field vanishes for $\kappa$ larger than a few units.}
\label{hstark}
\end{figure}
\end{center}

\begin{center}
\begin{figure}
\caption{The zfc and fc magnetization versus temperature curves are shown here for several  applied fields, ranging from $1.0 \;$kOe to $9.0 \;$kOe every $1.0 \;$kOe.
Below  $4.0 \;$kOe the curves can be considered as reversibles.}
\label{expmt}
\end{figure}
\end{center}

\begin{center}
\begin{figure}
\caption{ Isothermal magnetization cycles measured in  ascending (ZFC) and  descending (FC) fields are shown, starting from $3.5 \; $K to $8 \; $K every $0.5\;$K. The overlap of the inital part of the cycles into a single straight line defines the demagnetization factor of the sample.}
\label{expmh}
\end{figure}
\end{center}

\begin{center}
\begin{figure}
\caption{A selected isothermal magnetization cycle is plotted here both in the traditional and in the new ways, for the temperature of  $4.5 \;$K.
The field $H^{*}$ is hardly noticeable whereas the peak effect is the absolute minimum of the $2dMB$ vs $H$ curve.}
\label{exp2dmbha}
\end{figure}
\end{center}

\begin{center}
\begin{figure}
\caption{A selected isothermal magnetization cycle is plotted here both in the traditional and in the new ways, for the temperature of  $6.0 \;$K.
The field $H^{*}$ is the absolute minimum, whereas the peak effect is just a local one, for the $2dMB$ vs $H$ curve. 
}
\label{exp2dmbhb}
\end{figure}
\end{center}

\begin{center}
\begin{figure}
\caption{A selected isothermal magnetization cycle is plotted here both in the traditional and in the new ways, for the temperature of  $8.0 \;$K.
The field $H^{*}$ is the absolute minimum of the $2dMB$ vs $H$ curve, which is essentially totally reversible.}
\label{exp2dmbhc}
\end{figure}
\end{center}

\begin{center}
\begin{figure}
\caption{ This plot shows the theoretical curves ($\kappa = 3,\; 4\;  \mbox{and} \; 5$) 
and  that, among them, the $\kappa=4$ curve provides the best fitting to the $T = \; 8$K experimental curve.
All four curves were normalized to their maximum absoulte value. Along the $H$ axis the three theoretical curves were scaled by their $H_{c2}$, and the experimental curve was scaled such that its $H^{*}$ field is 0.46 away from the theoretical normalized $H_{c2}$ field.}
\label{comp}
\end{figure}
\end{center}

\begin{center}
\begin{figure}
\caption{The phase diagram of the Nb sample, containing its several critical fields, is shown here.}
\label{expht}
\end{figure}
\end{center}

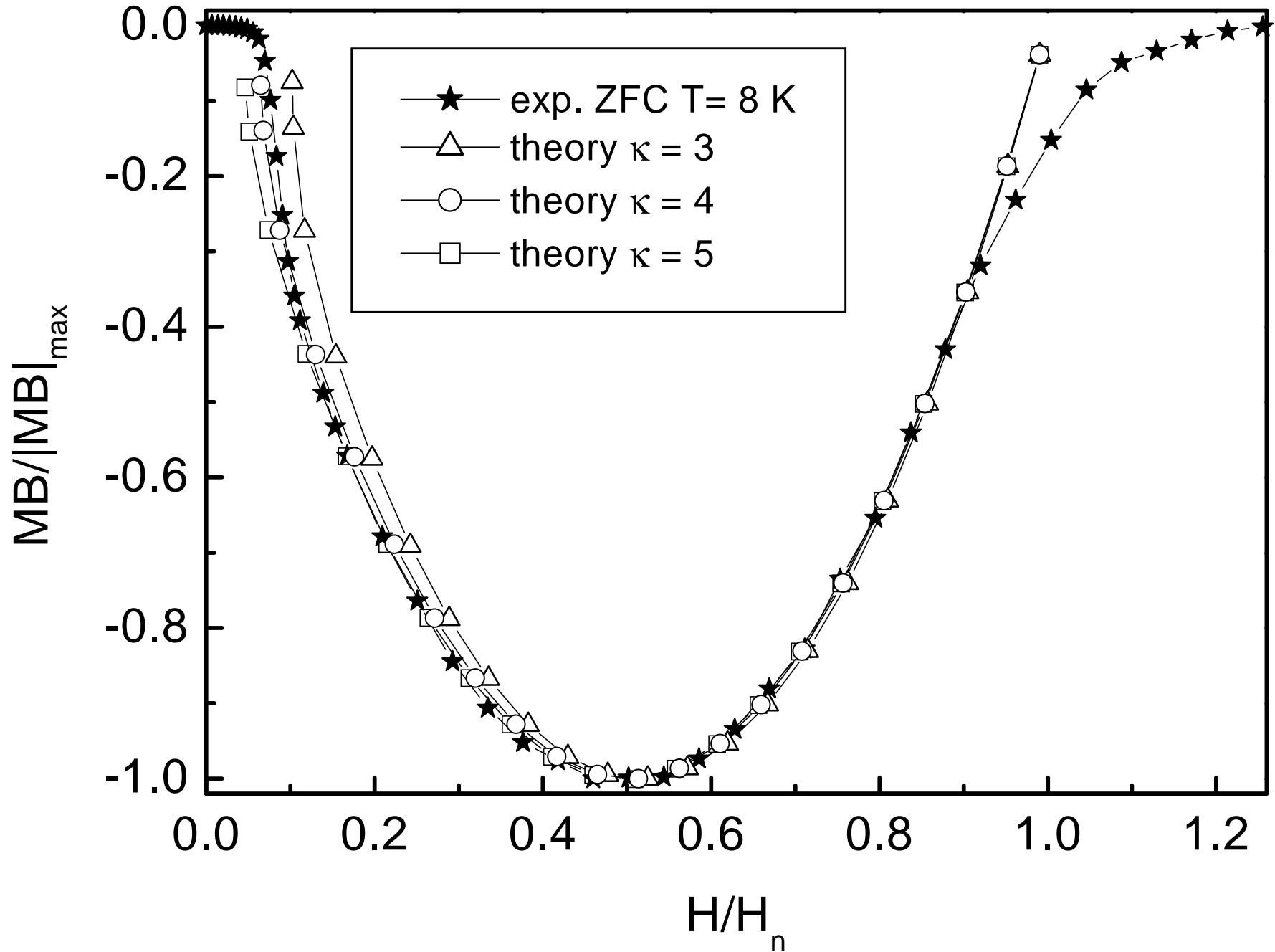
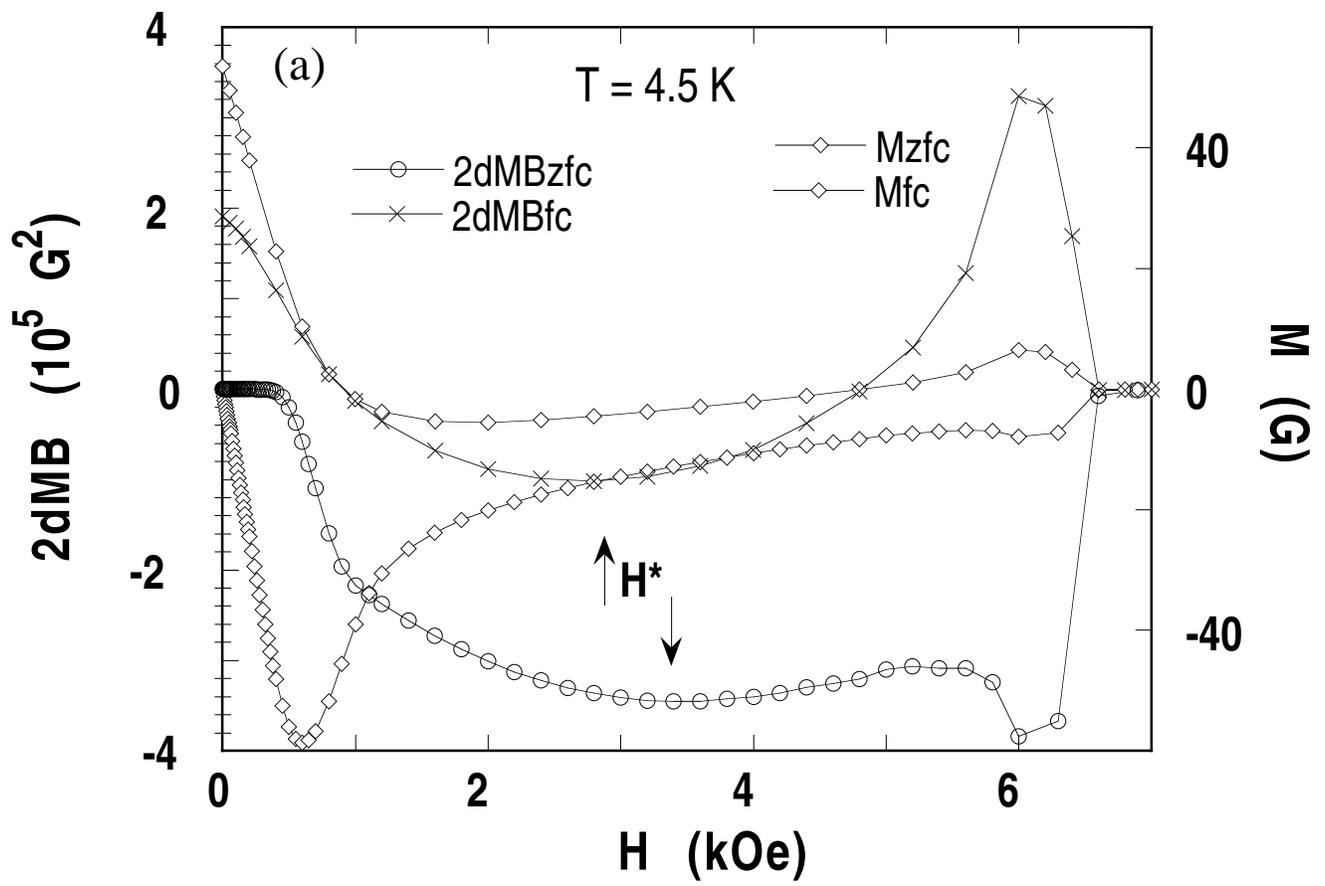
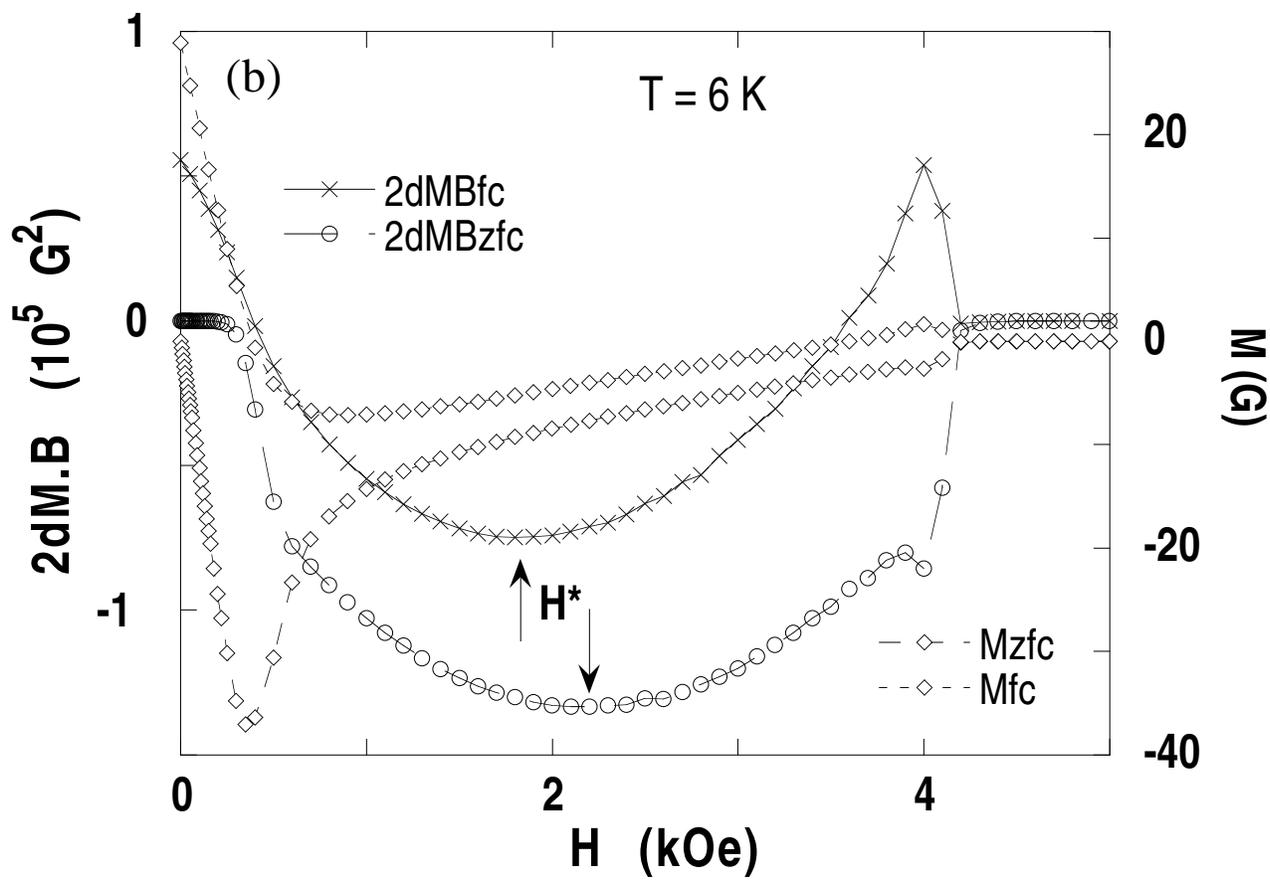
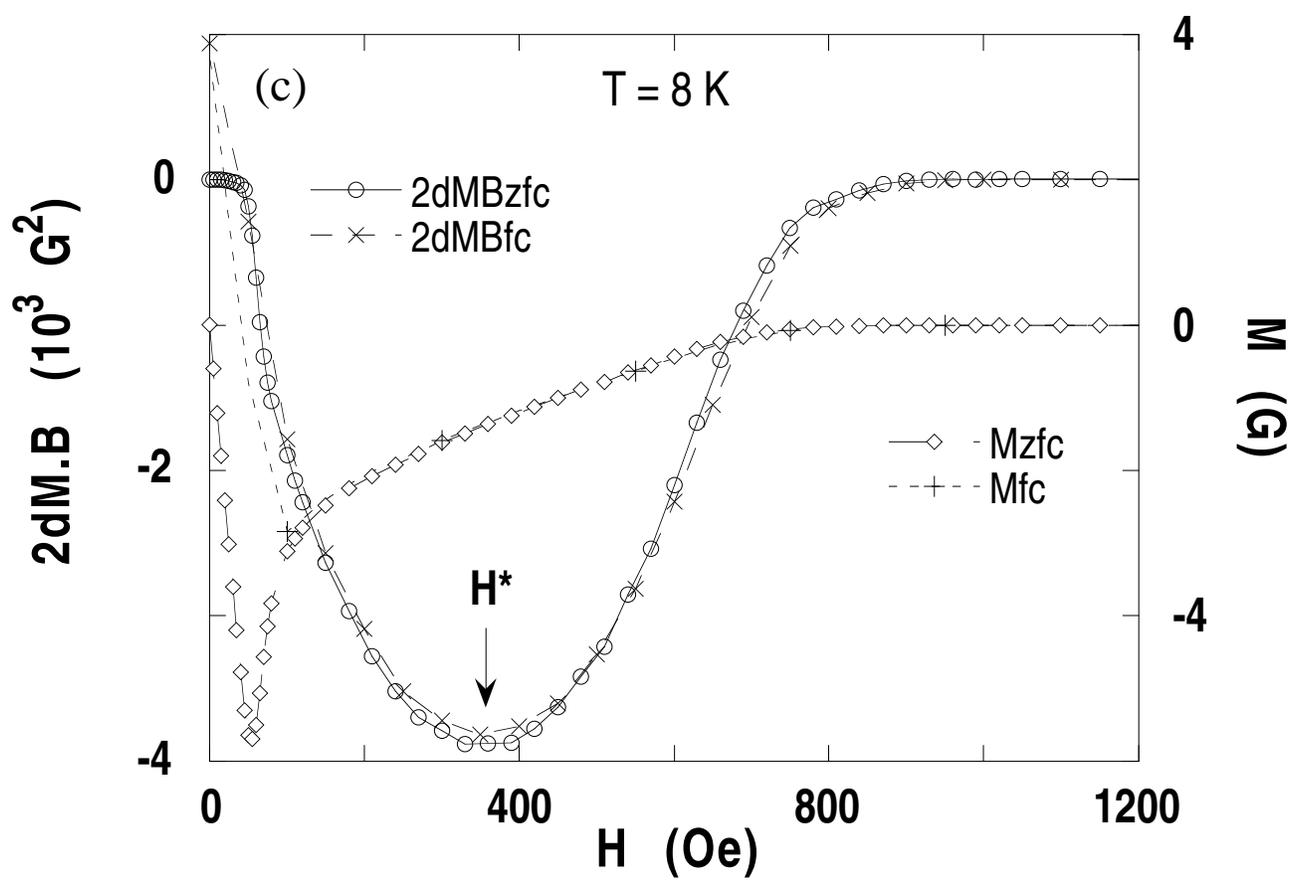
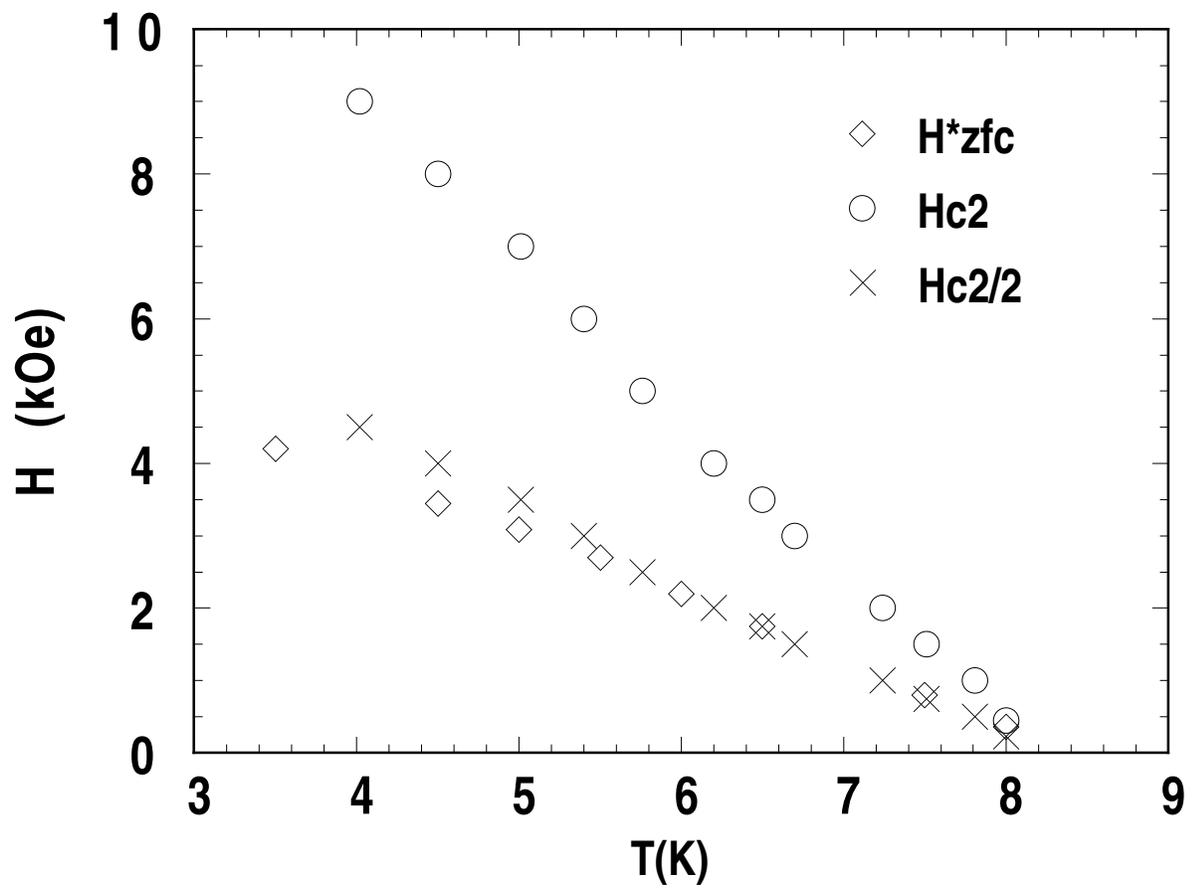
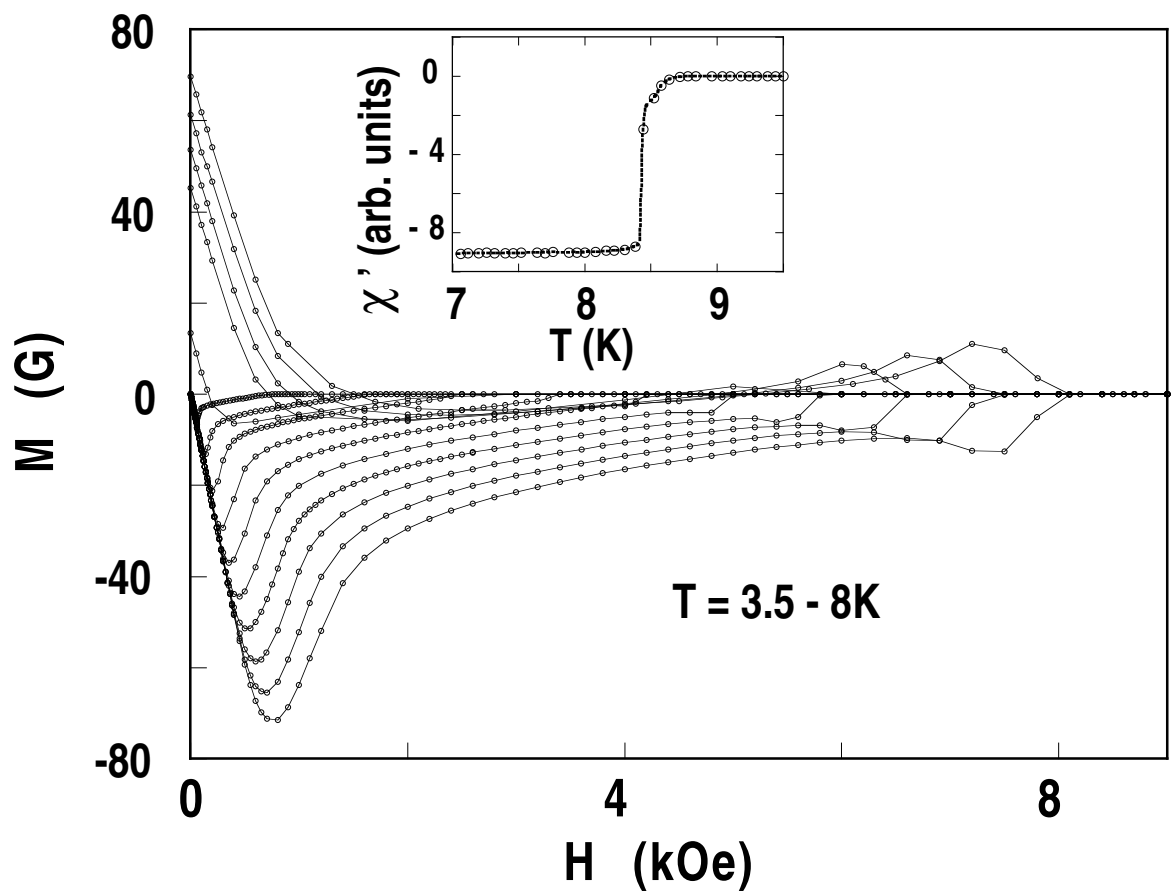
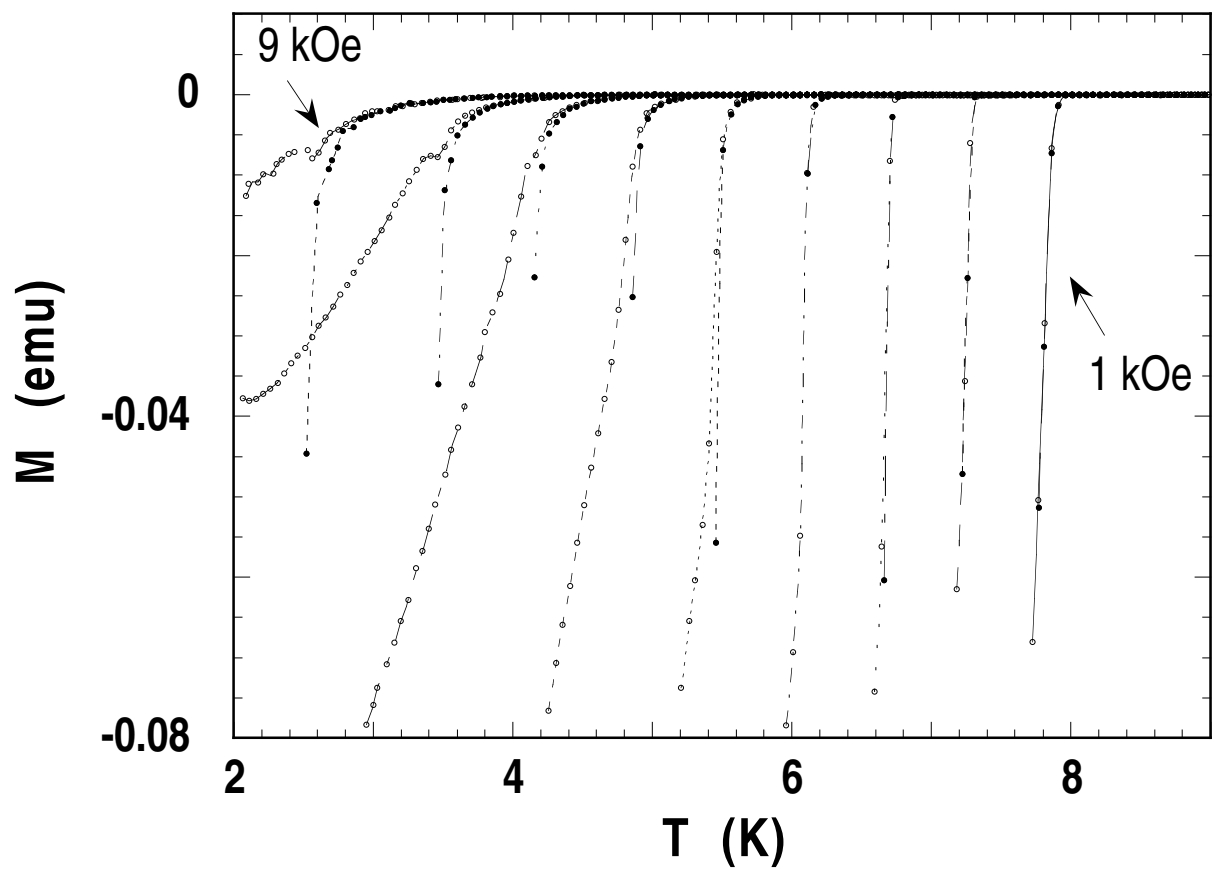
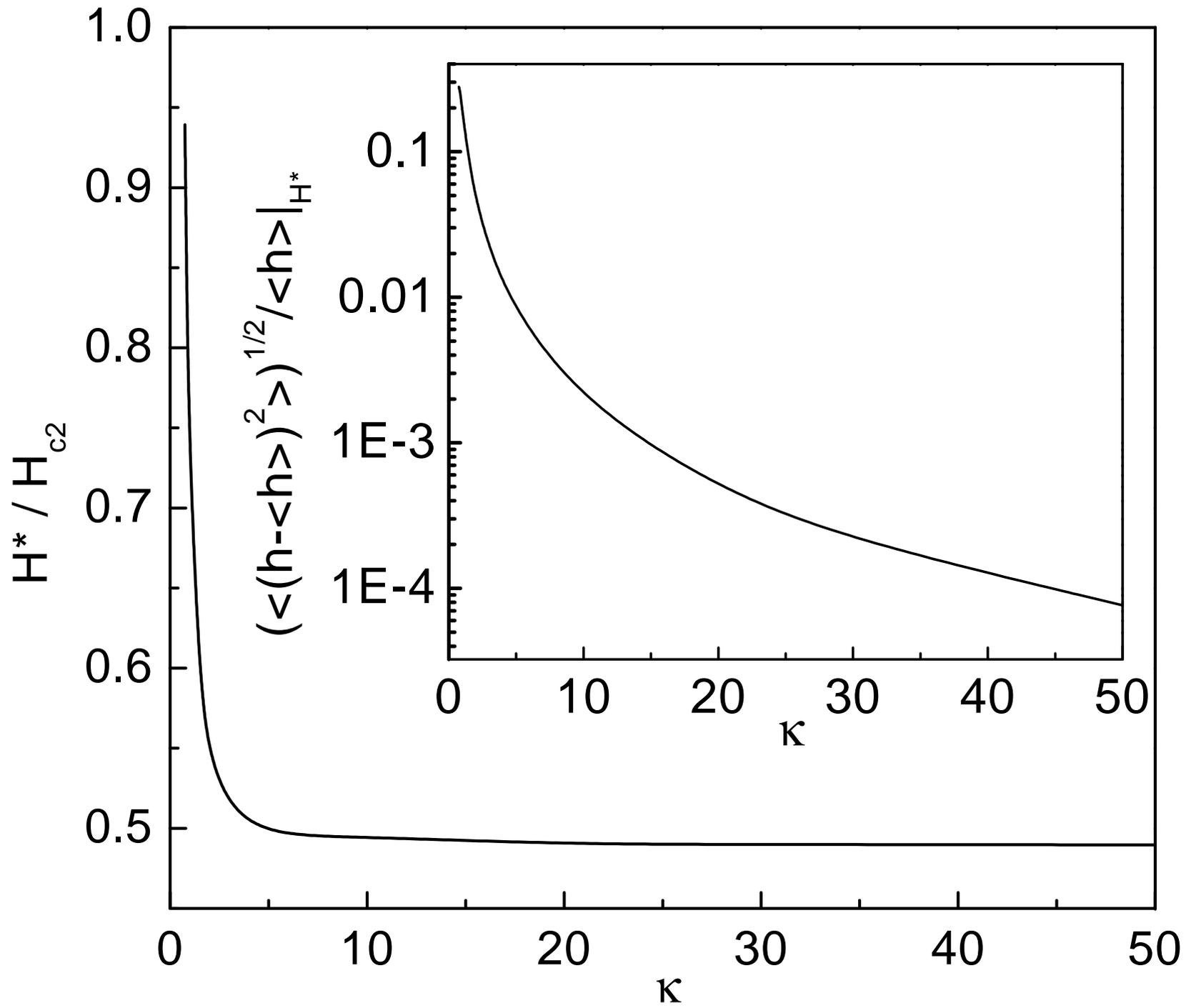
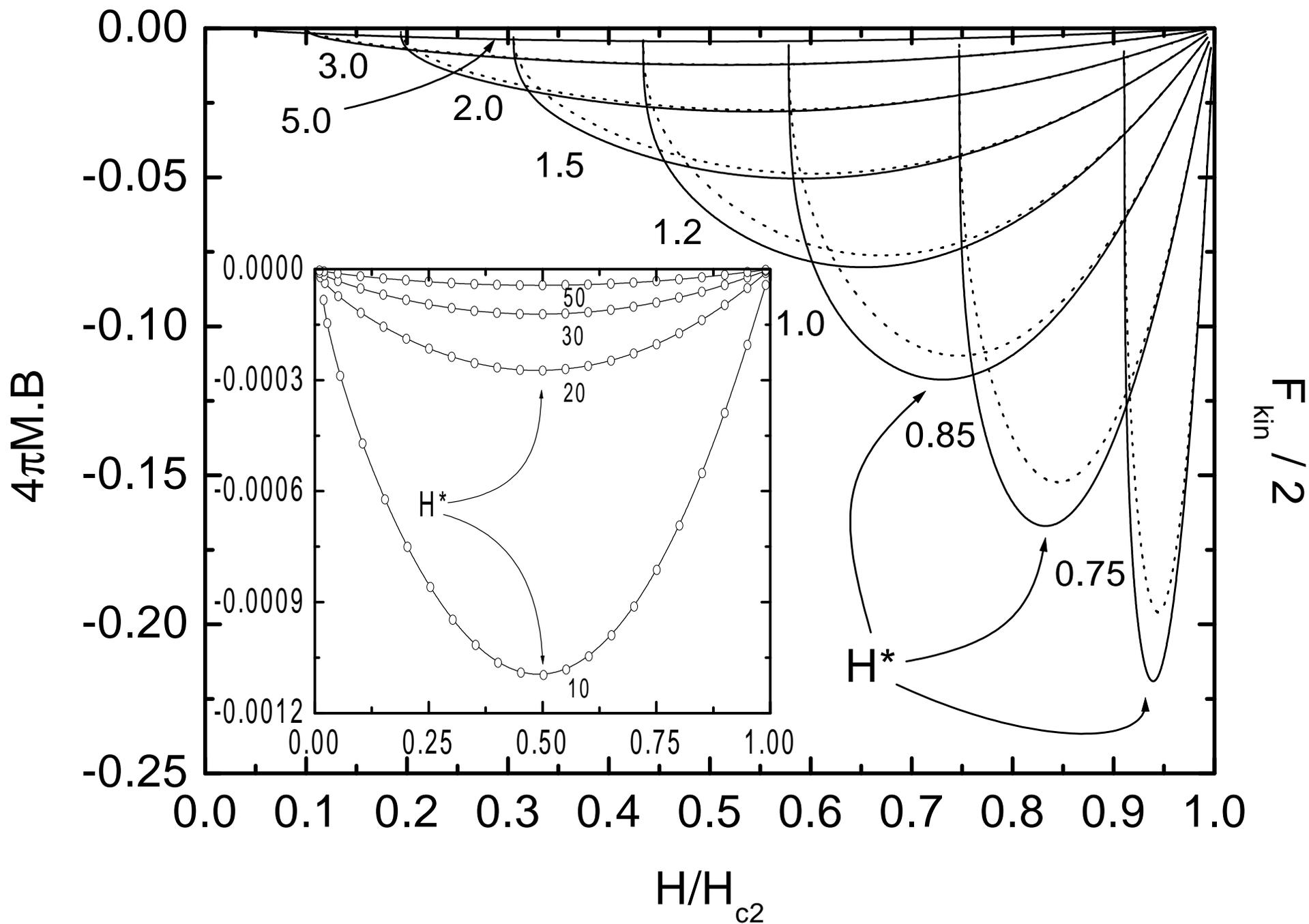
\end{document}